\documentclass[man, floatsintext]{apa7} 
\usepackage[utf8]{inputenc}
\usepackage[style=apa,sortcites=true,sorting=nyt,backend=biber]{biblatex}
\usepackage[american]{babel} 		
\usepackage{caption}				
\usepackage{amsmath}				
\usepackage{breqn}					
\usepackage[T1]{fontenc}			
\usepackage{amssymb}				
\usepackage{txfonts}				
\usepackage{rotating}				
\usepackage[utf8]{inputenc} 		
\usepackage{csquotes}
\usepackage{enumitem}
\usepackage{blindtext}
\usepackage{pdflscape}
\usepackage{color}
\usepackage{booktabs}
\usepackage{multirow}
\usepackage[LGR,T1]{fontenc} 		
\usepackage{pdfpages}
\usepackage{import}
\usepackage{threeparttable} 
\usepackage{titletoc}
\usepackage[]{hyperref}
\hypersetup{colorlinks=true,allcolors=black}
\usepackage{subcaption}

\DeclareMathOperator{\SD}{SD}

\DeclareLanguageMapping{english}{american-apa}

\AtEveryBibitem{\clearfield{number}}
\AtEveryBibitem{\clearfield{url}}
\AtEveryBibitem{\clearfield{urldate}}

\abstract{In this paper, we demonstrate a purely Bayesian approach for estimating within-group and between-group effect sizes for learning outcomes encountered in educational research, taking naturally into account the multilevel structure of the data, as well as heterogeneous residual variances among time points and conditions. We provide a detailed implementation using the brms package in R serving as a wrapper for the probabilistic programming language Stan. We recommend that for a pooled design, one computes an effect size $d_s$ similar to a Cohen's $d$, and for a paired design, one should compute two possibly different quantities $d_s$ and $d_z$ to correct for correlations in within-group designs and allow for comparability across different studies. All these effect sizes are based on ideas coming from Hedge's total effect size $\delta_t$ introduced in 2007. Ultimately, these estimates allow us to study the differential effectiveness of educational interventions with respect to classes.}


\bibliography{bibliography}
\usepackage{listings}

\title{A Bayesian Approach to Estimating Effect Sizes in Educational Research}

\shorttitle{Bayesian Effect Sizes in Educational Research}  

\authorsnames{Yannis Bähni} 

\authorsaffiliations{}

\DeclareMathOperator{\diff}{diff}

\DeclareMathOperator{\paired}{paired}

\DeclareMathOperator{\sd}{sd}
\DeclareMathOperator{\gain}{gain}
\DeclareMathOperator{\pr}{pr}
\DeclareMathOperator{\po}{po}
\DeclareMathOperator{\id}{id}

\begin{document}
\maketitle

\section{Methodological Implications Statement}
This article makes the estimation and interpretation of total variance effect sizes in multilevel modeling accessible and transparent for practitioners in educational research by providing a tutorial. Moreover, we present clear guidelines on when and how to use the different effect sizes belonging to the $d$-family in a realistic scenario. The originality is that, apart from reporting only an average effect size, one can compute the standard deviation of this estimate with respect to the clustering, providing a much more refined picture of the differential effectiveness of an educational intervention.

\section{Introduction}
Consider two groups of sizes $n_1$ and $n_2$, respectively, with means $\mu_1$ and $\mu_2$, and standard deviations $\sigma_1$ and $\sigma_2$. To answer the question of whether there is a statistically robust difference between a psychometric measure for the two groups, one can compute a standardized mean difference for the sample following (\cite[Equation~(1)]{lakens:d:2013}), by
\begin{equation}
    \label{eq:d_s}
    d_s := \frac{\mu_2 - \mu_1}{\sqrt{\frac{(n_1 - 1)\sigma_1^2 + (n_2 - 1)\sigma^2_2}{n_1 + n_2 - 2}}} = \sqrt{n_1 + n_2 - 2}\frac{\mu_2 - \mu_1}{\sqrt{(n_1 - 1)\sigma_1^2 + (n_2 - 1)\sigma^2_2}},
\end{equation}
which can be thought of as the mean difference divided by a weighted standard deviation. This standardized mean difference can be used to compare effects between empirical studies in meta-analyzes, as precise measures and scaling do not matter. However, such meta-analyzes in psychological sciences often overestimate the true effect size due to publication bias, as demonstrated in (\cite{wagenmakers:meta:2023}); that is, many publications only mention results in favor of a hypothesis, while results indicating no effect or even a negative one are neglected. In teaching and learning science, a typical setup consists of a pretest-posttest design, where the prior knowledge of a control and intervention condition is assessed before a particular teaching unit, and the learning achievements are measured by an identical posttest after the instruction. The effect size $d_s$ can be used to report differences in achievement between groups (pooled design) in either the pretest or the posttest, or to report learning gains from the pretest to the posttest in a repeated-measures design in a single group (within-group differences or paired design). In the first case, there are two different methods for computing this effect size. For example, as in Student's $t$-test, where one assumes equal variances (homoscedasticity assumption) and that the data are normally distributed, or akin to Welch's $t$-test, one can consider heterogeneous residual variances of the groups instead. Following (\cite{delacre:ttest:2017}), one should always use heterogeneous residual variances, and we fully support this since the point estimates for pooled ($\sigma_1 = \sigma_2$) and heterogeneous variances ($\sigma_1 \neq \sigma_2$) might not coincide in general.

In the case of a within-group or paired design, we have that $n_1 = n_2 = n$, and thus the standardized mean difference $d_s$ in \eqref{eq:d_s} reduces to
\begin{equation}
    \label{eq:d_s_paired}
    d_s = \frac{\mu_2 - \mu_1}{\sqrt{\frac{(n - 1)\sigma_1^2 + (n - 1)\sigma^2_2}{2n - 2}}} = \frac{\mu_2 - \mu_1}{\sqrt{\frac{\sigma_1^2 + \sigma^2_2}{2}}} = \sqrt{2}\frac{\mu_2 - \mu_1}{\sqrt{\sigma_1^2 + \sigma^2_2}},
\end{equation}
\noindent independent of the total sample size $n$, where the labels now stand for the first and second time points, respectively. As explained in (\cite{lakens:d:2013}), there are at least three more effect sizes belonging to the $d$-family to measure the learning gain in a paired sample design. Usually, they differ in how the difference between the means at different time points is divided by a suitable weighted interaction of the standard deviations of these observations. Unfortunately, researchers refer to an effect size as Cohen's $d$, regardless of how it was calculated. This can be biased and random, as results might not be reproducible or comparable. Hence, one should always use a subscript for an effect size belonging to the $d$-family when reporting results and provide the precise formula used to compute it. One of the main issues in computing effect sizes for a within-group design is the fact that the measures could be correlated, and the paired $d_s$ in \eqref{eq:d_s_paired} might overestimate the true effect size (\cite{dunlap:paired:1996}). As suggested in the paper (\cite{lakens:d:2013}), an effect size belonging to the $d$-family that takes the correlation between measurements into account is called Cohen's $d_z$, and is defined by
\begin{equation}
    \label{eq:d_z}
    d_z := \frac{\mu_2 - \mu_1}{\sigma_{\diff}},
\end{equation}
\noindent where $\sigma_{\diff}$ denotes the standard deviation of the difference between the means $\mu_2 - \mu_1$ with respect to time in the same group. This is similar to a paired sample $t$-test because, instead of two distributions for the pretest and posttest separately, we consider the raw learning gain given by the individual posttest solution rate minus the pretest solution rate.

The effect size $d_z$ allows for negative learning gains. As argued in (\cite{coletta:gain:2020}), this does not make much sense from a theoretical perspective because the lowest learning gain should be zero. However, negative learning gains make sense from a theoretical perspective in our view because cognitively activating instruction could initially confuse the students, causing them to perform worse on the posttest than on the pretest, or for other reasons. Therefore, we propose the following effect size for within-group designs
\begin{equation}
    \label{eq:paired}
    d_{\paired} := \{d_s, d_z\}.
\end{equation}
\noindent This means that for a within-group design, one computes two possibly different estimates of effect sizes for Cohen's $d$, and $d_z$ can additionally be computed for the normalized learning gain, as discussed in (\cite{schalk:prior_knowledge:2022}).

A major flaw of all the effect sizes discussed so far is that they are point estimates that could differ drastically from one another. One way to fix this issue is to use confidence intervals provided by statistical software. However, these intervals are based on bootstrapping mechanisms and are thus only approximations, assuming normal distributions. A much more reliable way to calculate the entire effect size distribution is provided by Bayesian statistics (\cite{schoot:bs:2021}). Here, one obtains full posterior distributions from the prior distribution as well as the likelihood function of the observations. This approach was prominently featured in (\cite{Kruschke:ttest:2012}) for effect sizes and is increasingly used in the evaluation of outcomes in empirical studies in psychology and the social sciences; see (\cite{peter:wd:2024}). The purpose of this article is to provide a detailed estimation of effect sizes belonging to the $d$-family, as mentioned above, and their implementation in the R-package brms, based on the statistical programming language Stan (\cite{buerkner:brms:2017}). For a nice introduction to the rudiments of this package, see (\cite{buerkner:intro:2019}). In contrast to the Bayesian effect sizes discussed in (\cite{Kruschke:ttest:2012}), we emphasize the inclusion of the multilevel structure and the use of heterogeneous residual variances. Moreover, including the multilevel structure allows us to estimate effect sizes with respect to classes and in contrast to (\cite{hedges:delta_t:2007}), we do not assume that both groups follow a normal distribution with equal variances. To conclude the introduction, we list some advantages and disadvantages of this approach to answer the question of why we prefer the Bayesian approach to effect sizes over the classical frequentist one. The advantages of the Bayesian approach are as follows:

\begin{enumerate}
    \item Bayesian estimation directly indicates the degree to which an intervention was effective by examining the highest probability density intervals instead of relying on $p$-values.
    \item One obtains a complete distribution of effect sizes along with credible intervals instead of a point-estimate and bootstrapped confidence intervals, implying that the estimated effect size lies with some fixed probability in a certain range given the data. In classical frequentist statistics, parameters are fixed population characteristics and not random variables.
    \item One can naturally take into account the multilevel structure of the data, as reviewed in (\cite{zitzmann:multilevel:2022}). Multilevel structured data naturally arise in realistic classroom settings in empirical studies where students are nested within classes. Usually, studies follow a within-class or between-class randomization, in which the control and intervention conditions are randomly distributed within or between classes.
    \item In a Bayesian framework, one can naturally estimate heterogeneous residual variances, allowing for heteroscedastic comparisons instead of only homogeneous ones. The heteroscedasticity assumption is much more natural in the teaching and learning sciences since the performance of different groups might vary a-priori.
    \item One does not have to assume that both groups are sampled from populations that follow a normal distribution, as assumed in Student's or Welch's $t$-test, by incorporating different modeling distributions. For example, choosing Student's $t$-distributions with an identity link leads to robust linear regression that is less influenced by outliers or skewed normal distributions, which corrects for the skewness of the data.
    \item Classical $t$-tests are only applicable to two groups. The Bayesian framework offers a much more flexible approach by allowing for multiple groups and time points, remaining robust over varying group sizes as well as different metrics for measuring learning gains.
\end{enumerate}
Some drawbacks, especially for researchers who are not familiar with the Bayesian approach and its implementation, are as follows:
\begin{enumerate}
    \item Bayesian models are based on stochastic sampling procedures with suitably chosen initial conditions. In some cases, the use of default settings might lead to biased estimates and convergence issues. To avoid this, we recommend the procedures outlined in (\cite{Kruschke:barg:2021}) to become more familiar with the Bayesian workflow and its obstacles.
    \item Bayesian estimates are robust but more time-consuming than classical methods, especially with larger sample sizes and when fitting complex models.   
    \item Effect sizes still depend heavily on the method chosen and serve only as rough indicators of the presence of an effect and its magnitude (\cite{kraft:d:2020}). Effect sizes should always be accompanied by a multitude of other statistical analyzes (\cite{ly:d:2021}) and should be interpreted carefully (\cite{bakker:effect:2019}). 
    \item There are some issues in treating hypothesis testing within the Bayesian framework and the classical $t$-value of a $t$-test. For an overview, see (\cite{paemel:bf:2010}; \cite{ly:bf:2016}; \cite{wong:ttest:2025}). However, we do not pursue this direction further in this article.
    \item Proper sensitivity analysis is required to assess the stability and robustness of the computed effect sizes. This should also be done in the frequentist approach to effect sizes, as especially in paired designs, there are quite a few different effect sizes (\cite{lakens:d:2013}).
\end{enumerate}

The article is structured as follows. First, we provide a brief introduction to the basics of multilevel modeling in brms with heterogeneous residual variances and their importance, highlighting key features of the data set in (\cite{peter:wd:2024}) available in the online material under \href{https://doi.org/10.17605/OSF.IO/8M3PJ}{https://doi.org/10.17605/OSF.IO/8M3PJ}. Then, we estimate and explore different effect sizes in the Bayesian framework for this data set and compare them with classical computations. Moreover, we explain how the inclusion of the multilevel structure can help quantify the differential effectiveness of educational interventions. Finally, we outline some guidelines for practitioners and discuss future methodological research in this direction.

\begin{figure}[h!tb]
    \caption{Distributions of Solution Rates in a Pretest-Posttest Design With Respect to Condition}
    \includegraphics[width=\textwidth]{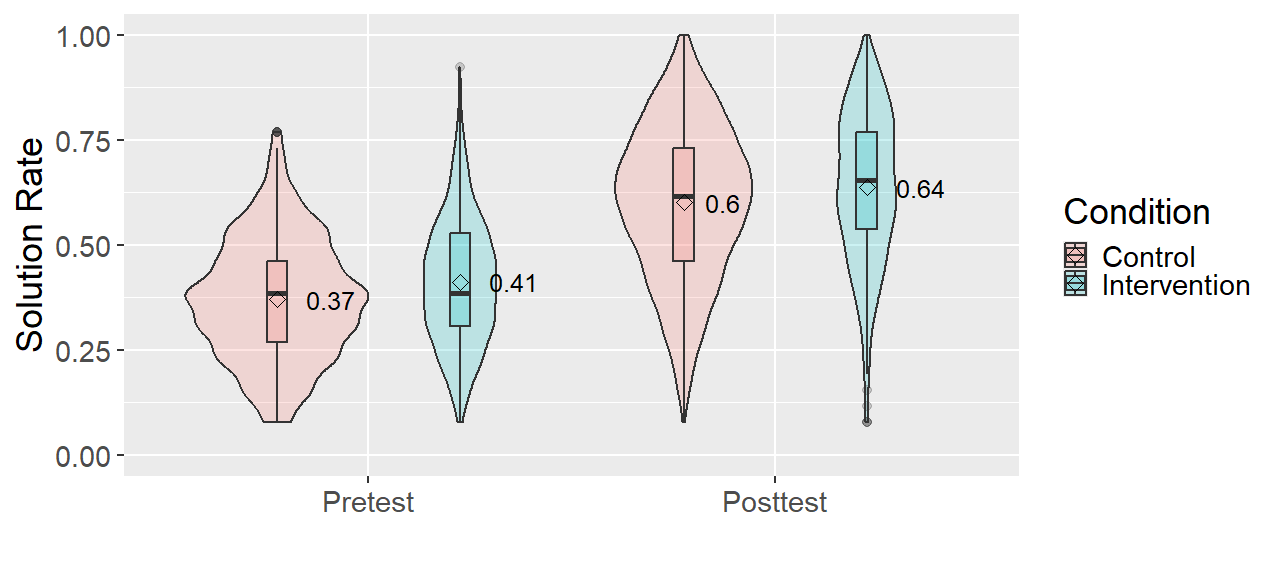}
\figurenote{Violin shapes indicate densities of score distributions, overlaid with box plots. Points above and below distributions indicate outliers. Squared points represent mean values with value indicated.}
    \label{fig:violin}
\end{figure}

\section{Technical Aspects of the Package brms}
Every linear multilevel model consists of predicting a dependent response variable $y$ by a certain distribution via a linear predictor variable $X\beta + Zu$. Here $X$ and $Z$ are $k \times l$-matrices, called design matrices; $\beta$ and $u$ are vectors of the same length $l$ as $y$, called population-level and group-level coefficients, respectively. The aim is to estimate the fixed effects $\beta$ and the random effects $u$, as well as additional model parameters determining the distribution based on the data provided in $y$ and $X,Z$. To showcase the R-package (\cite{R}), we use a real-life data set of a quasirandomized within-classroom pretest-posttest design (\cite{peter:wd:2024}) involving $n = 1377$ students divided into a control condition ($n_1 = 939$) and an intervention condition ($n_2 = 438$). All students received some teacher-guided lessons, but the two conditions differed in their predispositions. The aim is to provide an introduction to the Bayesian workflow. A much more detailed treatment is anticipated in the book in preparation (\cite{buerkner:brms:2024}). First, when dealing with unknown data, it is reasonable to gather some basic descriptive statistics. We assume that the data are tidy; that is, there are no incorrect or implausible entries. This is usually one of the most time-consuming steps in data analysis and is omitted here on purpose. Moreover, we do not address missing data imputation. Figure \ref{fig:violin} shows the complete distributions of the solution rates with respect to time and condition. One immediately sees that there is a constant effect of the intervention condition, which means that this condition outperforms the control condition at both time points to a similar extent. Quantifying how these observations persist when including the multilevel structure is the aim of a Bayesian regression model.

\subsection{A Bayesian Repeated-Measures ANOVA Model}
The first model is akin to a repeated-measures ANOVA model (\cite{peter:wd:2024}), incorporating two categorical variables: time (pretest vs. posttest) and condition (control vs. intervention), predicting the total solution rate (score) as the dependent variable; that is, a vector containing both the pretest and posttest solution rates. We deliberately use default improper flat prior distributions in this simple model, as specifying prior distributions would add another hurdle to using Bayesian modeling.

\begin{lstlisting}[language=R, caption=Bayesian repeated-measure model akin to an ANOVA model regressing the combined pretest and posttest scores on the interaction of time and condition, label = lst:ANOVA_1]
ANOVA_1 <- brm(bf(score ~ 0 + time * cond + (1|id)), 
  data = dat_long, 
  family = student())
\end{lstlisting}

\begin{figure}[h!tb]
    \caption{Graphical Posterior Predictive Check for the Posterior Distribution of the Score Generated by the Bayesian ANOVA Model Without Heterogeneous Residual Variances}
        \includegraphics[width=.85\textwidth]{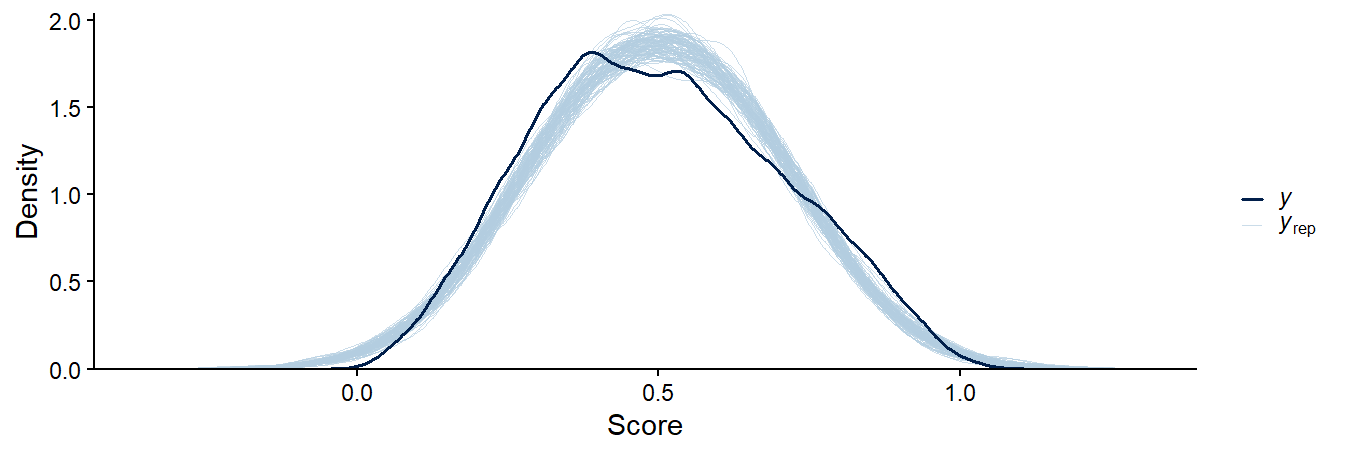}
\figurenote{$y =$ smoothed density function of the observed variable. $y_{\operatorname{rep}} = $ replication of the observed variable $y$ from draws of the posterior distribution.}
\label{fig:pp_check_ANOVA_1}
\end{figure}

The output of this model is presented in Table \ref{tab:ANOVA_1}. The results are directly interpretable by considering Figure \ref{fig:violin}, except for the logarithm of the residual standard deviation $\sigma$. The positive estimate of the intervention condition shows that it robustly outperforms the control condition in both the pretest and the posttest to a similar extent. This is indicated by the corresponding $95\%$ credible interval (CrI), which does not include zero. It should be remarked that $95\%$-CrIs are standard, but one should also consider $90\%$-CrIs, as their limits are usually estimated more reliably (\cite{kruschke:CrI:2018}; \cite{peter:wd:2024}). The $90\%$-CrI estimates can be found in the online material. The interaction term Posttest $\times$ Intervention shows that the advantage of the intervention condition remains constant in the posttest. To assess whether this model reasonably fits the data, we perform posterior predictive checks. The default graphical posterior predictive check for the model is shown in Figure \ref{fig:pp_check_ANOVA_1} and indicates that the model in Table \ref{tab:ANOVA_1} can be slightly improved. Indeed, the posterior distribution is skewed, whereas the Bayesian model does not appear to capture this. One possible solution is to use a skewed normal distribution in the family specification instead of an outlier-robust Student-$t$ distribution. However, this does not improve the graphical posterior plot in Figure \ref{fig:pp_check_ANOVA_1}, as one can check.

\begin{table}
  \begin{threeparttable}
    \caption{Results From a Bayesian Repeated-Measures ANOVA Model Regressing Solution Rates on Time, Condition, and Their Interaction}
    \label{tab:ANOVA_1}
    \begin{tabular}{@{}lccr@{}}         
    \toprule
         Parameter & Estimate & Error & $95\%$-CrI (HPD)\\\midrule
         Pretest (Control) & 0.37 & 0.01 & [0.36, 0.38]\\
         Posttest (Control) & 0.60 & 0.01 & [0.59, 0.61]\\
         Intervention & 0.04 & 0.01 & [0.02, 0.06]\\
         Posttest $\times$ Intervention & 0.00 & 0.01 & [-0.03, 0.02]\\
         $\log(\sigma)$ & 0.13 & 0.00 & [0.12, 0.13]\\
         $\sd_{\id}$ & 0.10 & 0.00 & [0.09, 0.11] \\\midrule
    \end{tabular}
    \tablenote{Error denotes the posterior standard deviation of the estimate (comparable to standard error). CrI = credible interval (highest probability density). $\sigma = $ residual standard deviation. $\sd_{\id} =$ standard deviation of multilevel hyperparameter for $\id$.}
  \end{threeparttable}
\end{table}

\subsection{Heterogeneous Residual Variances}
\begin{figure}[h!tb]
    \caption{Graphical Posterior Predictive Check for the Posterior Distribution of the Score Generated by the Bayesian ANOVA Model With Heterogeneous Residual Variances}
        \includegraphics[width=.9\textwidth]{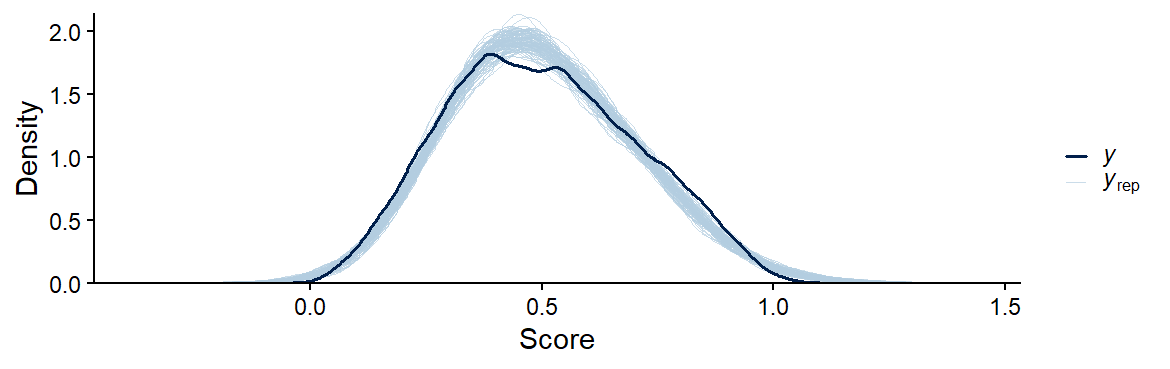}
\figurenote{$y =$ smoothed density function of the observed variable. $y_{\operatorname{rep}} = $ replication of the observed variable $y$ from draws of the posterior distribution.}
\label{fig:pp_check_ANOVA_2}
\end{figure}

To improve our model in Listing \ref{lst:ANOVA_1}, we add heterogeneous residual variances. Indeed, the model above assumes homogeneous residual variances among time and conditions, which might not be the case. In general, one should check this assumption prior to model fitting using the general model checks provided by the R-package performance (\cite{performance}). The improved model is described below in Listing \ref{lst:ANOVA_2}, and the output is presented in Table \ref{tab:ANOVA_2}. The posterior predictive check in Figure \ref{fig:pp_check_ANOVA_2} indicates that we implemented a reasonably fitting model that improves upon the one in Listing \ref{lst:ANOVA_1} and demonstrates the importance of considering heterogeneous residual variances to capture the skewness of the posterior distribution.

\begin{lstlisting}[language=R, caption=Bayesian repeated-measure model akin to an ANOVA model regressing the pretest and posttest scores on the interaction of time and condition with heterogeneous residual variances, label = lst:ANOVA_2]
ANOVA_2 <- brm(
    bf(score ~ 0 + time * condition + (1|id), 
    sigma ~ 0 + time * condition), 
    data = dat_long, family = student())
\end{lstlisting}

\begin{figure}[h!tb]
    \caption{Plot of the Mean Residual Error of the Score Generated by the Bayesian ANOVA Model With Heterogeneous Residual Variances}
        \includegraphics[width=\textwidth]{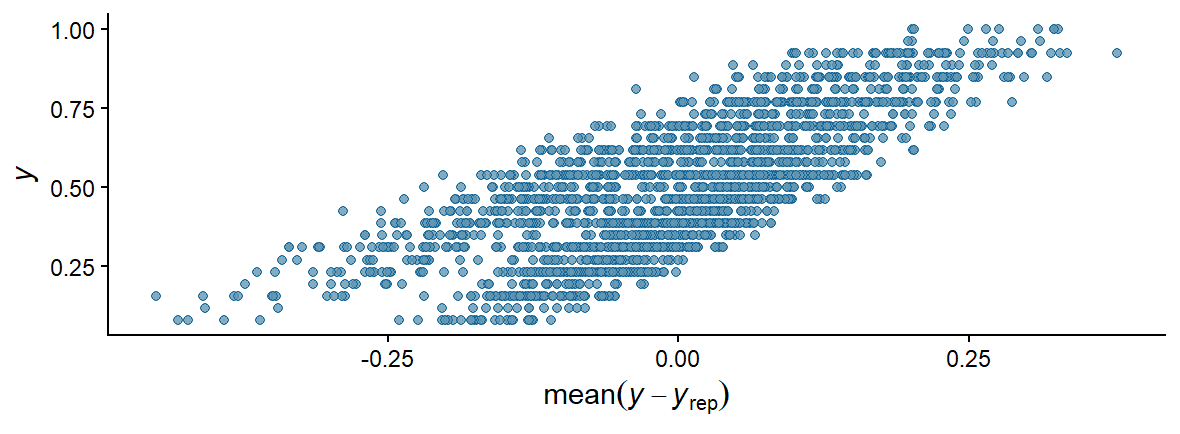}
\figurenote{$y =$ values of the observed variable. $y_{\operatorname{rep}} = $ replication of the observed variable $y$ from draws of the posterior distribution.}
\label{fig:pp_check_ANOVA_2_error}
\end{figure}

Apart from the standard convergence measures available in the online material, such as the R-hat estimate (for convergence, this value should ideally be $1$), the bulk and tail effective sample sizes (ESS), and the absence of divergent transitions, there is another informative graphical posterior check available. Figure \ref{fig:pp_check_ANOVA_2_error} shows the residual error plot of the model with heterogeneous residual variances. Ideally, this plot should be a point cloud without any visible patterns. As there is a linear pattern present, we might further improve our model.

\begin{table}
  \begin{threeparttable}
    \caption{Results From a Bayesian Repeated-Measures ANOVA
Model With Heterogeneous Residual Variances Regressing Solution Rates on Time, Condition, and
Their Interaction}
    \label{tab:ANOVA_2}
    \begin{tabular}{@{}lccr@{}}         
    \toprule
         Parameter & Estimate & Error & $95\%$-CrI (HPD)\\\midrule
         Pretest (Control) & 0.37 & 0.00 & [0.36, 0.38]\\
         Posttest (Control) & 0.60 & 0.01 & [0.59, 0.61]\\
         Intervention & 0.04 & 0.01 & [0.02, 0.06]\\
         Posttest $\times$ Intervention & -0.01 & 0.01 & [-0.03, 0.02]\\
         $\log(\sigma_{\operatorname{Pretest}})$ & -2.27 & 0.04 & [-2.36, -2.19]\\
         $\log(\sigma_{\operatorname{Posttest}})$ & -1.88 & 0.03 & [-1.94, -1.82]\\
         $\log(\sigma_{\operatorname{Intervention}})$ & 0.03 & 0.07 & [-0.10, 0.15]\\
         $\log(\sigma_{\text{Posttest:Intervention}})$ & -0.06 & 0.09 & [-0.24, 0.11]\\
         $\sd_{\id}$ & 0.10 & 0.00 & [0.09, 0.11]\\\midrule
    \end{tabular}
    \tablenote{Error denotes the posterior standard deviation of the estimate (comparable to standard error). CrI = credible interval (highest probability density). $\sigma = $ residual standard deviation. $\sd_{\id} =$ standard deviation of multilevel hyperparameter for $\id$.}
  \end{threeparttable}
\end{table}

\subsection{Random Intercepts and Random Slopes}
Usually, students are nested within classes, which gives a natural multilevel structure to the data. In our case, the class variable contains the necessary information about class membership ($62$ classes). To assess whether one should include this dependence in subsequent models, one traditionally computes an intraclass correlation coefficient (ICC). The code in Listing \ref{lst:icc} gives an estimated variance ratio of $0.17$ with a $95\%$-CrI of $[0.06, 0.26]$, which is comparable to a classical ICC. This means that approximately $17\%$ of the variance in the posttest solution rate can be attributed to systematic differences between the learners' classrooms, justifying multilevel modeling (\cite{zitzmann:multilevel:2022}). However, we advocate including the multilevel structure, irrespective of the Bayesian ICC value, as this provides valuable information about the deviation of estimates across classes (it is also a finding when there is no systematic variance arising from the clustering in classes). This can sometimes make model estimation slow and result in convergence issues. Especially if one wants to include random slopes and intercepts for residual standard deviations.

To incorporate the multilevel structure of the data provided by the nesting in classes, we add random intercepts and random slopes for the interaction term time $\times$ condition over class to the model in Listing \ref{lst:ANOVA_2}. We note that including random slopes over the condition is only possible because the study follows a quasirandomized within-classroom design. Moreover, this can naturally be done for the residual standard deviation $\sigma$ as well. The altered model specification is presented in Listing \ref{lst:ANOVA_3}. The sampling requires considerable time due to the complexity of the model, and we provide additional model specifications to improve runtime. In particular, we include normally distributed prior distributions for the parameters, as opposed to the improper flat priors used by default. To compare model performance, we used leave-one-out (LOO) cross-validation (\cite{loo:2017}). The model with normal prior distributions had an expected log pointwise predictive density difference of $-1.6$ with a standard error of $1.6$ compared to the model with default improper flat prior distributions. Consequently, no model substantially outperformed the others in terms of predictive accuracy, implying that the model is not sensitive to the choice of priors. This can additionally be checked by comparing the parameter estimates between the two models. The LOO-method can also be used to systematically study the effects of simplifying the model in Listing \ref{lst:ANOVA_3} by removing random effects from the residual variance or by using another posterior distribution family. For the LOO-method, models need not be nested.

The model output in Table \ref{tab:ANOVA_3} now contains the standard deviations of parameters over the random effects compared to Table \ref{tab:ANOVA_2}. This will be needed to compute effect sizes while taking into account the variance between classes. The graphical posterior predictive check for the residual error shows only a marginal improvement when including the multilevel structure compared to Figure \ref{fig:pp_check_ANOVA_2_error}. This is because our model contains very few predictors that do not explain additional systematic variances of the score. For example, there could be an influence of sex (female vs. male) or cognitive ability. As we are not interested in finding an optimal model but rather in a sufficiently good model that allows us to estimate effect sizes, we do not investigate this further. The search for an optimal model is discussed, for example, in (\cite{buerkner:brms:2024}). To further assure model fit quality, grouped graphical posterior plots (time and condition), as well as additional graphical calibration diagnostics, are included in the online material (ECDF and PIT-ECDF).

\begin{lstlisting}[language=R, caption=Estimating a Bayesian ICC value, label = lst:icc]
icc <- brm(
  bf(po ~ (1|class)), 
  data = dat, family = skew_normal())
performance::variance_decomposition(icc, ci = 0.95)
\end{lstlisting}

\begin{lstlisting}[language=R, caption=Bayesian repeated-measure multilevel model akin to an ANOVA regressing the combined pretest and posttest scores on time and condition with heterogeneous residual variances and including random intercepts and random slopes over classes, label = lst:ANOVA_3]
ANOVA_3 <- brm(
  bf(score ~ 0 + time * cond + (1|id) + (0 + time * cond|class), 
     sigma ~ 0 + time * cond + (0 + time * cond|class)), 
  data = dat_long, family = student(),
  prior = c(prior(normal(0, 1), class = "b"),
    prior(normal(-1, 2), class = "b", dpar = "sigma")),
  warmup = 2500, iter = 5000, cores = parallel::detectCores(),
  control = list(adapt_delta = 0.95, max_treedepth = 15))
\end{lstlisting}

\begin{table}
  \begin{threeparttable}
    \caption{Results From a Bayesian Repeated-Measures ANOVA
Model With Heterogeneous Residual Variances including Random Intercepts and Random Slopes over Classes Regressing Solution Rates on Time, Condition, and
Their Interaction}
    \label{tab:ANOVA_3}
    \begin{tabular}{@{}llcr@{}}         
    \toprule
         Parameter & Estimate(SD) & Error & $95\%$-CrI(HPD)\\\midrule
         Pretest (Control) & 0.36(0.05) & 0.01 & [0.35, 0.38]\\
         Posttest (Control) & 0.59(0.09) & 0.01 & [0.56, 0.61]\\
         Intervention & 0.04(0.02) & 0.01 & [0.01, 0.06]\\
         Posttest $\times$ Intervention & 0.00(0.04) & 0.02 & [-0.03, 0.03]\\
         $\log(\sigma_{\operatorname{Pretest}})$ & -2.35(0.09) & 0.05 & [-2.45, -2.26]\\
         $\log(\sigma_{\operatorname{Posttest}})$ & -2.04(0.09) & 0.04 & [-2.11, -1.96]\\
         $\log(\sigma_{\operatorname{Intervention}})$ & 0.03(0.10) & 0.08 & [-0.12, 0.18]\\
         $\log(\sigma_{\text{Posttest:Intervention}})$ & -0.04(0.09) & 0.10 & [-0.23, 0.16]\\
         $\sd_{\id}$ & 0.09 & 0.00 & [0.09, 0.10]\\\midrule
    \end{tabular}
    \tablenote{SD of the estimate denotes the standard deviation over classes. Error denotes the posterior standard deviation of the estimate (comparable to standard error). CrI = credible interval (highest probability density).}
  \end{threeparttable}
\end{table}

\section{Between-Group Differences (Pooled Design)}
\subsection{The Effect Size $d_s$}
As is apparent from Figure \ref{fig:violin} and the results in Table \ref{tab:ANOVA_3}, the intervention condition outperforms the control condition at both time points to a similar extent. An immediate question is what the magnitude of this effect is; that is, its effect sizes. More importantly, we would also like to know how homogeneous this effect is with respect to the different classes (\cite{edelsbrunner:lpa:2025}). Using the Bayesian framework, we can answer this question with a suitable model by fixing the time and letting the condition vary in the repeated-measures multilevel ANOVA model provided in Listing \ref{lst:ANOVA_3}. This is conceptually similar to Welch's $t$-test, as we consider unequal variances; however, it is not an exact replacement since we consider a Bayesian multilevel regression model with group-specific residual standard deviations. The output of this model is presented in Table \ref{tab:ttest_pr}. The control condition is considered the first group, and the intervention condition is considered the second. In contrast to the model in Listing \ref{lst:ANOVA_3}, we do not remove the intercept; thus, the estimated parameter for the intervention condition directly represents the mean difference $\mu_2 - \mu_1$ between the two conditions.

\begin{lstlisting}[language=R, caption=A Bayesian multilevel version of Welch's two-sample $t$-test between conditions, label = lst:ttest_pr]
ttest_pr <- brm(bf(
  pr ~ cond + (0 + cond|class),
  sigma ~ 0 + cond + (0 + cond|class)),
  family = student(), data = dat,
  prior = c(
    prior(normal(0, 1), class = "b"),
    prior(normal(-1, 2), class = "b", dpar = "sigma")),
  warmup = 2500, iter = 5000, cores = parallel::detectCores(),
  control = list(adapt_delta = 0.95, max_treedepth = 15))
\end{lstlisting}

\begin{figure}[h!tb]
    \caption{Histogram of Posterior Samples of the Effect Size $d_s$}
        \includegraphics[width=.9\textwidth]{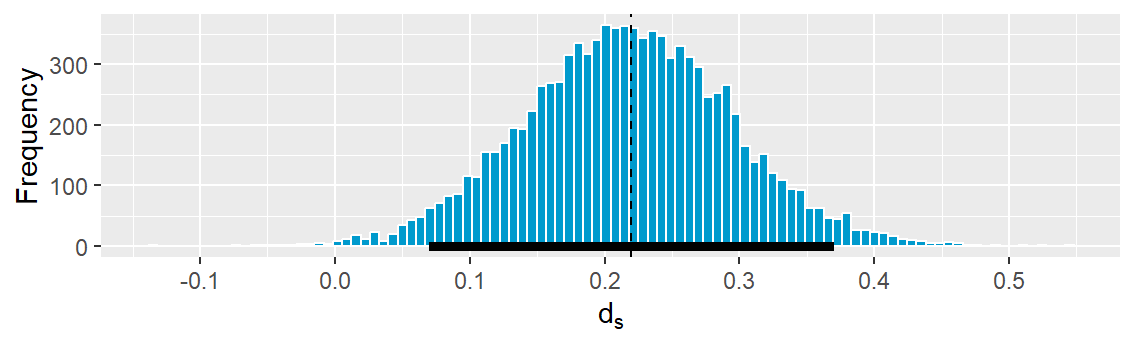}
\figurenote{Frequency describes the frequency over all four Markov chains. The black horizontal line represents the $95\%$-CrI. The dashed vertical line represents the mean value.}
\label{fig:d_s_distribution}
\end{figure}

\begin{figure}[h!tb]
    \caption{Differential Effectiveness of the Educational Intervention With Respect to Time and Classes}
        \includegraphics[width=\textwidth]{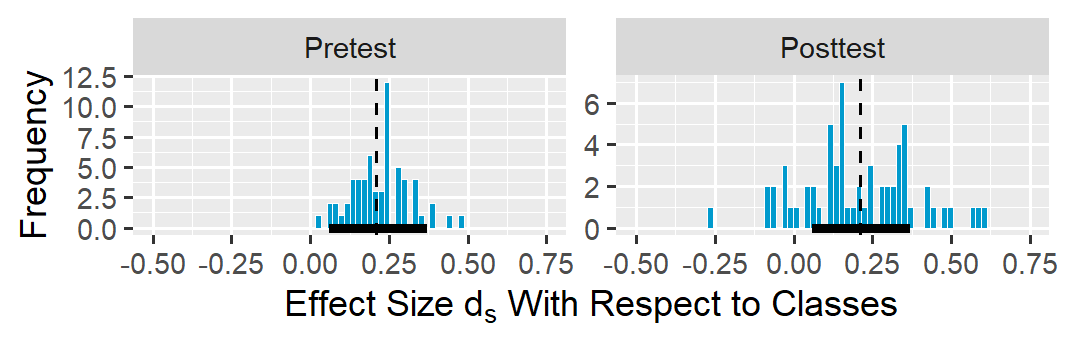}
\figurenote{Frequency describes the frequency of $d_s$ estimates with respect to classes. The black horizontal line represents the $95\%$-CrI of the mean $d_s$ estimates with respect to time. The dashed vertical line represents the mean values of $d_s$ with respect to time.}
\label{fig:d_s_cl}
\end{figure}

Using the output in Table \ref{tab:ttest_pr}, we can include fixed effects and standard deviations of the random effects in the computation of the effect size $d_s$ in \eqref{eq:d_s}. Following (\cite{hedges:delta_t:2007}; \cite{d_t:2019}), we capture all the effects arising from the multilevel structure using the total variance formula
\begin{equation*}
    d_s = \sqrt{n_1 + n_2 - 2}\frac{\mu_2 - \mu_1}{\sqrt{(n_1 - 1)(\sigma_1^2 + \sd_1^2 + \sd_{\sigma_1}^2) + (n_2 - 1)(\sigma^2_2 + \sd_2^2 + \sd_{\sigma_2}^2)}}.
\end{equation*}
This approach captures all sources of variation in the described multilevel model. Conceptually, it is given by the estimated difference between the group means divided by the square root of the sum of all variance components, in analogy to \eqref{eq:d_s}. By using the values up to maximal precision and not the rounded ones in Table \ref{tab:ttest_pr}, we obtain $d_s = 0.22$. Having the full posterior distribution of all parameters at hand, we can compute the full posterior distribution of the $d_s$-effect size. If we use the above formula for all four Markov chains, we obtain a $95\%$-CrI of $[0.07, 0.37]$ and the posterior distribution for $d_s$ depicted in Figure \ref{fig:d_s_distribution}. Since this credible interval does not include zero, we conclude that there is a statistically robust difference between the control and intervention conditions for the pretest. This value should be compared with $d_s = 0.28$ using a bootstrapped $95\%$ confidence interval (CI) of $[0.17, 0.40]$, without utilizing the multilevel structure of the data from a classical Cohen's $d$ computation using the R-package effectsize (\cite{effectsize}). In the posttest, we compute the effect size $d_s = 0.21$ with a $95\%$-CrI $[0.05, 0.37]$ in the same manner. This should again be compared with $d_s = 0.20$ ($95\%$-CI $[0.08, 0.31]$) without taking into account the multilevel structure. Therefore, at the posttest, the Bayesian and classical effect sizes $d_s$ are almost identical. However, we claim that the Bayesian version provides us with much more information. Indeed, since our computations involve the random structure, we can estimate effect sizes with respect to each class. Therefore, we obtain a standard deviation of $0.09$ of $d_s$ at the pretest across classes, and a standard deviation of $0.18$ of $d_s$ at the posttest. The corresponding distributions of the class-wise effect sizes $d_s$ are depicted in Figure \ref{fig:d_s_cl}. One immediately sees the differential effectiveness of the intervention with respect to classes. At the pretest, the advantage of the intervention condition compared to the control condition was rather homogeneous in terms of positive effect sizes with respect to classes. However, at the posttest, the range of possible effect sizes widened, with some individuals in the intervention condition apparently benefiting more than others. For example, in some classes, the control condition outperformed the intervention condition, as indicated by a negative estimate for the class-wise $d_s$.

\begin{table}
  \begin{threeparttable}
    \caption{Results From a Bayesian Multilevel Version of a Two-Samples $t$-Test With Heterogeneous Residual Variances Comparing Solution Rates Between Conditions at the Pretest}
    \label{tab:ttest_pr}
    \begin{tabular}{@{}lccr@{}}         
    \toprule
         Parameter & Estimate(SD) & Error & $90\%$-CrI (HPD)\\\midrule
         Intercept ($\mu_1$) & 0.36 & 0.01 & [0.35, 0.38]\\
         $\mu_2 - \mu_1$ & 0.04 & 0.01 & [0.01, 0.06]\\
         $\log(\sigma_1)$ & -2.02 & 0.03 & [-2.08, -1.97]\\
         $\log(\sigma_2)$ & -1.98 & 0.04 & [-2.07, -1.90]\\
         $\sd_1$ & 0.05 & 0.01 & [0.04, 0.07]\\
         $\sd_2$ & 0.05 & 0.01 & [0.03, 0.08]\\
         $\sd_{\sigma_1}$ & 0.06 & 0.03 & [0.00, 0.13]\\
         $\sd_{\sigma_2}$ & 0.08 & 0.06 & [0.00, 0.21]\\\midrule
    \end{tabular}
    \tablenote{Error denotes the posterior standard deviation of the estimate (comparable to standard error). CrI = credible interval (highest probability density). $\sigma = $ residual standard deviation. $\sd = $ standard deviation of multilevel hyperparameters.}
  \end{threeparttable}
\end{table}

\section{Within-Group Differences (Paired Design)}
\subsection{The Effect Size $d_s$}
In a paired design setup, we can compute the effect size $d_s$ from the Bayesian Multilevel model in Listing \ref{lst:ANOVA_3}, as shown in Listing \ref{lst:ANOVA_control}, by fixing the condition and allowing time to vary. The corresponding total variation effect size of \eqref{eq:d_s_paired} is then given by
\begin{equation*}
    d_s = \sqrt{2}\frac{\mu_{\po} - \mu_{\pr}}{\sqrt{\sigma_{\pr}^2 + \sd_{\pr}^2 + \sd_{\sigma_{\pr}}^2 + \sigma^2_{\po} + \sd_{\po}^2 + \sd_{\sigma_{\po}}^2 + 2 \sd_{\operatorname{\id}}^2}}.
\end{equation*}
\noindent Using this formula, for the control condition, we compute $d_s = 1.18$ with a $95\%$-CrI of $[0.88, 1.42]$ and a standard deviation of $0.41$ across classes. All computational details can be found in the online material. This estimate should be compared to $d_s = 1.40$ with a $95\%$-CI of $[1.33, 1.47]$, without taking into account the multilevel structure. Likewise, we performed the same computation for the intervention condition, resulting in $d_s = 1.23$ with a $95\%$-CrI of $[0.78, 1.58]$ and a standard deviation of $0.29$ across classes. This is comparable to the learning gain of the control condition, aligning with the observation that the $95\%$-CrI of the estimate of the interaction term Posttest $\times$ Intervention presented in Table \ref{tab:ANOVA_3} is zero; thus, both conditions gained similarly. Again, this should be compared to $d_s = 1.35$ with a $95\%$-CI of $[1.25, 1.45]$ using a classical computation. Moreover, we see that $d_s \pm \SD$ does not necessarily lie in the $95\%$-CrI already illustrated in Figure \ref{fig:d_s_cl}. This is due to the fact that the $95\%$-CrI only measures the uncertainty of the mean $d_s$ estimate over classes and might be small, especially for larger sample sizes. Therefore, the standard deviation of $d_s$ across the classes provides another informative measurement not captured by any standard metrics.

\begin{lstlisting}[language=R, caption=A Bayesian multilevel version of Welch's two-sample $t$-test within the control condition, label = lst:ANOVA_control]
ANOVA_control <- brm(bf(
    score ~ time + (1|id) + (0 + time|class), 
    sigma ~ 0 + time + (0 + time|class)), 
  data = dat_long %>% filter(cond == "Control"), 
  family = student(),
  prior = c(
    prior(normal(0, 1), class = "b"),
    prior(normal(-1, 2), class = "b", dpar = "sigma")
  ),
  warmup = 2500,
  iter = 5000,
  cores = parallel::detectCores(),
  control = list(adapt_delta = 0.95, max_treedepth = 15))
\end{lstlisting}

\begin{table}
  \begin{threeparttable}
    \caption{Results From a Bayesian Repeated-Measures ANOVA
Multilevel Model With Heterogeneous Residual Variances Regressing Solution Rates on Time in the Control Condition}
    \label{tab:ANOVA_control}
    \begin{tabular}{@{}lccr@{}}         
    \toprule
         Parameter & Estimate & Error & $95\%$-CrI (HPD)\\\midrule
         Intercept ($\mu_{\pr}$) & 0.37 & 0.01 & [0.35, 0.38]\\
         $\mu_{\po} - \mu_{\pr}$ & 0.23 & 0.01 & [0.20, 0.25]\\
         $\log(\sigma_{\pr})$ & -2.34 & 0.05 & [-2.45, -2.25]\\
         $\log(\sigma_{\po})$ & -2.04 & 0.04 & [-2.12, -1.97]\\
         $\sd_{\pr}$ & 0.05 & 0.01 & [0.04, 0.07]\\
         $\sd_{\po}$ & 0.09 & 0.01 & [0.07, 0.11]\\
         $\sd_{\sigma_{\pr}}$ & 0.09 & 0.06 & [0.00, 0.23]\\
         $\sd_{\sigma_{\po}}$ & 0.11 & 0.05 & [0.01, 0.22]\\
         $\sd_{\id}$ & 0.09 & 0.01 & [0.04, 0.07]\\\midrule
    \end{tabular}
    \tablenote{Error denotes the posterior standard deviation of the estimate (comparable to standard error). CrI = credible interval (highest probability density). $\sigma = $ residual standard deviation. $\sd = $ standard deviation of multilevel hyperparameters.}
  \end{threeparttable}
\end{table}

\subsection{The Effect Size $d_z$}
Both effect sizes $d_s$ for the control and intervention conditions possibly overestimate the true effect size, as the correlations $r = 0.34$ for the control condition's pretest and posttest, as well as the correlation $r = 0.40$ for the intervention condition, were not considered. Thus, we implement a multilevel model for the raw learning gain $\po - \pr$ in Listing \ref{lst:gain} to correct this. It is convenient to simultaneously model the raw learning gains of the control and intervention conditions to guarantee comparability between the two conditions. In contrast to the previous model, we need to convert the dataset from long to wide format for this computation. The results of this model are presented in Table \ref{tab:gain}.

\begin{lstlisting}[language=R, caption=A Bayesian multilevel version of the paired sample Welch's $t$-test, label = lst:gain]
ttest <- brm(bf(
    po - pr ~ 0 + cond + (0 + cond|class),
    sigma ~ 0 + cond + (0 + cond|class)),
  family = student(),
  data = dat,
  prior = c(
    prior(normal(0, 1), class = "b"),
    prior(normal(-1, 2), class = "b", dpar = "sigma")
  ),
  warmup = 2500, iter = 5000,
  cores = parallel::detectCores(),
  control = list(adapt_delta = 0.95, max_treedepth = 15))
\end{lstlisting}

By adapting formula \eqref{eq:d_z} and taking into account all the different sources of variance in the multilevel model, we obtain the effect size
\begin{equation}
    \label{eq:gain}
    d_z = \frac{\gain_i}{\sqrt{\sigma^2_i + \sd^2_i + \sd_{\sigma_i}^2}}, \qquad i = 1,2.
\end{equation}
\noindent For the control condition, we estimate $d_z = 1.11$ with a $95\%$-CrI of $[0.88, 1.30]$ and a standard deviation of $0.39$. This should be compared with $d_z = 1.21$ with a $95\%$-CI of $[1.13, 1.30]$ without including the multilevel structure of the data. For the intervention condition, we compute $d_z = 1.20$ with a $95\%$-CrI of $[0.91, 1.43]$ and a standard deviation of $0.30$. This should be compared with $d_z = 1.23$ ($95\%$-CI $[1.12, 1.33]$) without any multilevel structure. Comparing these estimates to those from the previous section, we see that indeed $d_z < d_s$; however, the difference is small. Finally, we remark that the computation of this model is much more efficient than estimating the complex model in Listing \ref{lst:ANOVA_control} twice . 

\begin{table}
  \begin{threeparttable}
    \caption{Results From a Bayesian Multilevel Version of a Paired Sample $t$-Test With Heterogeneous Residual Variances Comparing Solution Rates Between Conditions}
    \label{tab:gain}
    \begin{tabular}{@{}lccr@{}}         
    \toprule
         Parameter & Estimate & Error & $95\%$-CrI\\\midrule
         Control ($\gain_1$) & 0.22 & 0.01 & [0.20, 0.25]\\
         Intervention ($\gain_2$) & 0.23 & 0.02 & [0.20, 0.26]\\
         $\log(\sigma_1)$ & -1.82 & 0.03 & [-1.88, -1.76]\\
         $\log(\sigma_2)$ & -1.81 & 0.04 & [-1.89, -1.73]\\
         $\sd_1$ & 0.09 & 0.01 & [0.07, 0.12]\\
         $\sd_2$ & 0.08 & 0.01 & [0.05, 0.11]\\
         $\sd_{\sigma_1}$ & 0.08 & 0.04 & [0.00, 0.16]\\
         $\sd_{\sigma_2}$ & 0.07 & 0.05 & [0.00, 0.17]\\\midrule
    \end{tabular}
    \tablenote{Error indicates standard deviation of the estimate (comparable to
standard error). CI = credible interval. $\sigma = $ residual standard deviation. $\sd = $ standard deviation of multilevel hyperparameters.}
  \end{threeparttable}
\end{table}

\subsection{The Effect Size $d_n$}
In this final subsection, we consider normalized learning gains
\begin{equation*}
    \gain_n := \begin{cases}
        \displaystyle\frac{\po - \pr}{1 - \pr}, &\pr < 1,\\
        \po - 1, & \pr = 1, 
    \end{cases}
\end{equation*}
instead of the raw learning gains discussed in the previous section. We use this to showcase the flexibility of the Bayesian approach for estimating effect sizes. The implementation of the corresponding model is presented in Listing \ref{lst:gain_n}. We do not provide the output since the interpretation of the parameter estimates is not as straightforward as in the case of raw learning gains; however, we can compute $d_n = 1.04$ with a $95\%$-CrI of $[0.87, 1.21]$ and a standard deviation of $0.38$ across classes using \eqref{eq:gain}. For the intervention condition, we obtain $d_n = 1.07$ with a $95\%$-CrI of $[0.83, 1.27]$ and a standard deviation of $0.28$

\begin{lstlisting}[language=R, caption=A Bayesian multilevel version of the paired sample Welch's $t$-test using normalized learning gains instead of raw learning gains, label = lst:gain_n]
dat <- dat %>% mutate(
    gain_n = ifelse(pr < 1, (po - pr)/(1 - pr), po - pr))

ttest_n <- brm(bf(
  gain_n ~ 0 + cond + (0 + cond|class),
  sigma ~ 0 + cond + (0 + cond|class)),
  family = skew_normal(), data = dat,
  prior = c(
    prior(normal(0, 1), class = "b"),
    prior(normal(-1, 2), class = "b", dpar = "sigma")),
  warmup = 2500, iter = 5000,
  cores = parallel::detectCores(),
  control = list(adapt_delta = 0.95, max_treedepth = 15))
\end{lstlisting}

\section{Discussion}
In a realistic pretest-posttest setup with a control and intervention condition, we have seen how to associate different relative measures of learning gains and achievements belonging to the $d$-family, combining (\cite{lakens:d:2013}; \cite{Kruschke:ttest:2012}) and ideas from (\cite{hedges:delta_t:2007}; \cite{d_t:2019}). If we fix time, we can compare two (or more) conditions and calculate an effect size $d_s$ that measures a standardized difference between the groups, taking into account heterogeneous residual variances and the multilevel structure of the data often encountered in educational research. If one takes into account the multilevel structure of the data, the effect sizes $d_s$ and $d_z$ will often differ from a non-cluster-adjusted effect size. A difference is expected due to the different variance structure used in the first case, and many complex factors contribute to the estimation. However, for large sample sizes with a reasonable number of classes (a rule of thumb is at least thirty different classes), this structure must be included, as it provides a much more refined picture of the variances between classes through the standard deviation of the effect size. It makes sense from a theoretical perspective that the overall effect size differs if there is significant variance between classes, as there could be strong or weak classes. Simply computing average effect sizes without accounting for these differences might overestimate the true effect size. This is of particular importance in longitudinal field studies, where learning gains from classes are measured in a real-life environment and might exhibit strong between-classroom differences.

For within-group estimates of effect sizes, there are two viable candidates. First, the effect size $d_s$ behaves like the corresponding one for between-group designs. However, this quantity generally overestimates the true effect size, as the correlation of the measures at the two time points is not considered. Thus, a better choice is to use Cohen's $d_z$, which is also easier to compute in the Bayesian framework and should be reported for between-group designs since it provides a Bayesian version of the $t$-value using the formula $t = \sqrt{n}d_z$, where $n$ denotes the total sample size. As a general rule, one should calculate $d_s$ and $d_z$ carefully and report all of them together with their respective $95\%$-credible intervals, as suggested in \eqref{eq:paired}. Again, any analysis should rely on multiple statistical methods; however, communicating effect sizes is still important and should be done with the appropriate subscripts to make transparent which method was used to compute them. In addition, one could compute the effect size $d_n$ derived from normalized learning gains. This measure could be useful when a ceiling effect is present at the posttest. Finally, one should report Bayesian diagnostics, such as different graphical posterior predictive checks, convergence measures, and precise model specifications, including prior distributions and Markov chain Monte Carlo settings, in supplemental material. Performing such checks not only guaranties the reproducibility of findings but also improves the model specification used for estimating Bayesian multilevel effect sizes.

\subsection{Conclusions and Implications}
The derived guidelines for using the different Bayesian multilevel effect sizes are summarized in Table \ref{tab:guidelines}. Using the Bayesian framework, the information provided by the different effect sizes contributes to analyzing the differential effectiveness of educational interventions. Indeed, even if the mean effect size has a positive credible interval, its standard deviation across classes can still be large. Overall, this differential perspective should prevent the overestimation of effect sizes and make its interpretation more nuanced. Whenever conducting multilevel modeling, one should report the standard deviations of the estimates across the multilevel structure, as shown, for example, in Table \ref{tab:ANOVA_3} and Table \ref{tab:ttest_pr}.

\begin{table}
  \begin{threeparttable}
    \caption{Derived Guidelines for Estimating Bayesian Multilevel Effect Sizes}
    \label{tab:guidelines}
    \begin{tabular}{@{}llll@{}}         
    \toprule
     Randomization & Design & Effect Size & Remarks\\\midrule
     Within-Class & Paired & $d_s(\SD)$, $\underline{d_z(\SD)}$, $d_n(\SD)$ & Random intercept for condition;\\
     & Pooled & $\underline{d_s(\SD)}$ & Mean effect size; report $95\%$-CrI\\
     Between-Class & Paired & $d_s(\SD)$, $\underline{d_z(\SD)}$, $d_n(\SD)$ & No random intercept for condition;\\
     & Pooled & $\underline{d_s}$ & Mean effect size; report $95\%$-CrI\\\midrule
    \end{tabular}
    \tablenote{The underlined effect size is the preferred one.}
  \end{threeparttable}
\end{table}

\subsection{Limitations and Future Research}
Estimating Bayesian models is time consuming and presents many more hurdles than classical frequentist methods. Even for experts, settling on prior distributions, choosing the correct posterior distributions, and fine-tuning parameters is difficult. In particular, this is true for beginners. Methodologically, the current study should be supplemented with simulation studies testing the systematic effects of different clustering of the classes, the effects of varying residual variances, the dependence on distributions, and the influence of different correlations of the dependent variables on estimating Bayesian multilevel effect sizes. An aim for future research is also the systematic study of the performance of different effect sizes in the presence of floor or ceiling effects, as well as missing data. Moreover, a detailed analytical comparison of the effect sizes in (\cite{hedges:delta_t:2007}) and the ones used in this paper could be of value from a mathematical perspective. Lastly, although estimating effect sizes varies based on cluster membership, we are aware that multilevel modeling is still a variable-centered method (\cite{saqr:ai:2026}). In contrast to person-centered methods like latent profile analysis and mixture modeling, variable-centered methods estimate some kind of population grand-mean, and no heterogeneity between individuals is qualitatively or quantitatively captured.

\section{Code Availability}
The code and dataset used in this study are freely available on the Open Science Framework (OSF) under \href{https://doi.org/10.17605/OSF.IO/8M3PJ}{https://doi.org/10.17605/OSF.IO/8M3PJ}.
\printbibliography

\end{document}